\documentclass{article}

\usepackage{arxiv}

\usepackage[utf8]{inputenc}
\usepackage[T1]{fontenc}
\usepackage{hyperref}
\usepackage{url}
\usepackage{booktabs}
\usepackage{amsfonts}
\usepackage{nicefrac}
\usepackage{microtype}
\usepackage{graphicx}
\usepackage{float}
\usepackage{algorithm}
\usepackage{algpseudocode}
\graphicspath{ {./images/} }
\title{Five Fatal Assumptions: Why T-Shirt Sizing Systematically Fails for AI Projects}

\author{
 Raja Soundaramourty\textsuperscript{1*}, Ozkan Kilic\textsuperscript{1}, Ramu Chenchaiah\textsuperscript{1} \\
  \\
  \textsuperscript{1}Cisco Systems, Inc. \\
  \texttt{\{rajasoun, okilic, rchencha\}@cisco.com} \\
  \textsuperscript{*}Corresponding author \\
}

\begin{document}
\maketitle

\begin{abstract}
Agile estimation techniques, particularly T-shirt sizing, are widely used in software development for their simplicity and utility in scoping work. However, when we apply these methods to artificial intelligence initiatives---especially those involving large language models (LLMs) and multi-agent systems---the results can be systematically misleading. This paper shares an evidence-backed analysis of five foundational assumptions we often make during T-shirt sizing. While these assumptions usually hold true for traditional software, they tend to fail in AI contexts: (1) linear effort scaling, (2) repeatability from prior experience, (3) effort-duration fungibility, (4) task decomposability, and (5) deterministic completion criteria.

Drawing on recent research into multi-agent system failures \cite{cemri2025}, scaling principles \cite{kim2025}, and the inherent unreliability of multi-turn conversations \cite{laban2025}, we show how AI development breaks these rules. We see this through non-linear performance jumps, complex interaction surfaces, and ``tight coupling'' where a small change in data cascades through the entire stack. To help teams navigate this, we propose \textbf{Checkpoint Sizing}: a more human-centric, iterative approach that uses explicit decision gates where scope and feasibility are reassessed based on what we learn during development, rather than what we assumed at the start. This paper is intended for engineering managers, technical leads, and product owners responsible for planning and delivering AI initiatives.
\end{abstract}

\keywords{agile methodology \and AI development lifecycle \and effort estimation \and large language models \and multi-agent systems \and software project management \and T-shirt sizing}

\section{Introduction}

\subsection{Why Traditional Estimation Fails AI Teams}

T-shirt sizing---categorizing work items as Small (S), Medium (M), Large (L), or Extra-Large (XL)---has served software engineering well for decades. Its appeal lies in its simplicity: relative sizing avoids the false precision of hour-based estimates while providing a shared language for teams to plan their work. The method relies on a simple idea: if an experienced team has built something similar before, they can use that pattern to estimate the next project with reasonable confidence.

But when we bring this same mindset to AI, the estimates often fall apart. We aren't just talking about being off by a few days; we're seeing projects expand by months, or hit technical ``walls'' that weren't even on the radar. A task that looks like a ``chatbot'' on the surface can diverge wildly in effort once you start dealing with data quality, evaluation hurdles, and the unpredictable nature of multi-turn conversations \cite{laban2025, coelho2024}. In these cases, the ``slippage'' isn't usually a failure of the team---it's a property of the problem itself.

The critical insight is that these failures aren't about poor execution or lack of skill. It's that the mental model we use for traditional software just doesn't fit AI. T-shirt sizing rests on five implicit assumptions that we take for granted in software, but which collapse when they meet the unique, often non-linear world of AI development.

\subsection{Contributions}

This paper makes the following contributions:
\begin{enumerate}
    \item \textbf{Identification of Five Fatal Assumptions:} We systematically enumerate and analyze five foundational assumptions underlying T-shirt sizing that fail in AI contexts.
    \item \textbf{Empirical Grounding:} We ground each assumption's failure in recent empirical literature (including arXiv preprints and established software-engineering studies) on multi-agent system behavior, agent-system scaling properties, and AI development workflows.
    \item \textbf{Quantitative Evidence:} We provide concrete metrics demonstrating the magnitude of estimation errors, including $N(N-1)$ interaction complexity growth in multi-agent systems and a 39\% average performance degradation in multi-turn conversations \cite{laban2025}.
    \item \textbf{Alternative Framework:} We propose Checkpoint Sizing as an evidence-based alternative methodology specifically designed for AI project estimation.
\end{enumerate}

\subsection{Paper Organization}

The remainder of this paper is organized as follows. Section 2 reviews related work in agile estimation and AI project management. Section 3 describes our methodology for identifying and evaluating the five assumptions. Section 4 presents detailed analysis of each assumption's failure mode. Section 5 discusses practical implications and introduces the Checkpoint Sizing alternative. Section 6 concludes with directions for future work.

\subsection{Scope (What Counts as an ``AI Project'' Here)}

This paper uses ``AI projects'' to refer to initiatives where the primary delivery risk is driven by \textbf{model behavior and data-dependent uncertainty}, rather than deterministic code alone. This includes (i) \textbf{LLM applications} (prompted systems that must meet reliability and safety targets), (ii) \textbf{agentic workflows} (single- or multi-agent tool-using systems with coordination and verification overhead), (iii) \textbf{retrieval-augmented generation (RAG)} systems (where performance depends on corpus quality, retrieval, grounding, and evaluation), and (iv) \textbf{model adaptation} such as fine-tuning, steering, or continual improvement loops.

We also include classic ML when the initiative's critical path is dominated by dataset readiness, evaluation design, and performance/safety validation. Pure software work that happens to call an API is in-scope only when these AI-specific sources of uncertainty materially determine effort, schedule, or definition-of-done.

\section{Background and Related Work}

\subsection{Agile Estimation Methods}

Agile estimation encompasses several techniques designed to size work without false precision. Story points assign relative complexity values to user stories, typically using Fibonacci sequences (1, 2, 3, 5, 8, 13\ldots) to reflect estimation uncertainty at larger scales~\cite{cohn2005}. Planning poker leverages collective team wisdom through simultaneous card reveals to surface divergent assumptions~\cite{grenning2002}. T-shirt sizing simplifies further by mapping work to categorical buckets (S, M, L, XL) that correspond roughly to duration ranges.

These methods share common assumptions: that teams possess relevant prior experience, that similar-looking tasks require similar effort, and that work can be decomposed into parallelizable components. Cone of Uncertainty models acknowledge early-stage estimation variance but assume convergence as projects progress through defined phases~\cite{mcconnell2006}.

\subsection{AI Project Management Literature}

Recent studies have documented systematic challenges in AI project estimation. Amershi et al.~\cite{amershi2019} identified nine distinct characteristics of machine learning workflows that differ from traditional software, including data dependencies, experimental iteration, and model decay. Sculley et al.~\cite{sculley2015} characterized technical debt in ML systems, noting that conventional software engineering intuitions often lead practitioners astray.

The emergence of large language models and multi-agent architectures has introduced additional complexity dimensions. Research on multi-agent coordination failures~\cite{cemri2025} identifies 14 distinct failure modes clustered into system design issues, inter-agent misalignment, and task verification. Studies on agent-system scaling~\cite{kim2025} quantify non-linear performance regimes and coordination-dependent trade-offs that defy intuition trained on traditional software.

\subsection{Estimation Failures in LLM-Integrated Systems}

While outcome-level AI project failure rates are frequently discussed, this paper focuses on estimation failure mechanisms observable in the development process itself. Recent work highlights that including LLM-based intelligent interfaces can introduce estimation challenges that are not captured by traditional sizing approaches, and motivates richer specifications that account for data sources, interfaces, and algorithms~\cite{coelho2024}.

\section{Methodology}

\subsection{Assumption Identification}

We derived the five assumptions through a structured synthesis of how T-shirt sizing is used in practice and how its underlying reasoning maps to the constraints of AI development:

\begin{enumerate}
    \item \textbf{Estimation-method synthesis:} We reviewed canonical agile estimation approaches (e.g., story points, planning poker, relative sizing) to identify the implicit premises that make categorical sizing usable~\cite{cohn2005,grenning2002,mcconnell2006}.
    \item \textbf{AI workflow mismatch grounding:} We anchored those premises against established software-engineering characterizations of ML/AI workflows (e.g., data dependency, experimentation, and ML-specific technical debt)~\cite{amershi2019,sculley2015}.
    \item \textbf{Empirical anchoring in modern LLM systems:} We mapped each premise to empirical findings in LLM/agent literature, prioritizing work that quantifies failure modes, scaling behavior, and multi-turn unreliability~\cite{cemri2025,kim2025,laban2025,coelho2024}.
    \item \textbf{Assumption-to-impact mapping:} For each premise, we articulated the mechanism by which its violation produces estimation error (scope underestimation, integration underestimation, or schedule compression overestimation).
\end{enumerate}

\subsection{Evaluation Criteria}

Each assumption was evaluated against three criteria:

\begin{enumerate}
    \item \textbf{Validity in Traditional Software:} Does the assumption hold for conventional software development?
    \item \textbf{Violation in AI Contexts:} Is the assumption systematically violated by AI development characteristics?
    \item \textbf{Estimation Impact:} Does the violation produce materially incorrect estimates?
\end{enumerate}

Assumptions meeting all three criteria were classified as ``fatal''---foundational beliefs that cause systematic estimation failure when applied to AI projects.

\subsection{Evidence Collection}

We collected supporting evidence from empirical literature, with particular emphasis on recent arXiv publications addressing multi-agent systems, LLM behavior, and effort estimation in LLM-integrated software. Each cited work was evaluated for direct relevance to the identified assumptions (see Appendix~A for validation methodology).

\section{Analysis of the Five Fatal Assumptions}

\subsection{Assumption 1: Linear Effort Scaling}

\subsubsection{The Traditional Model: ``Double the Features, Double the Work''}

In traditional software, we've learned that effort usually scales in a straight line. An XL project is roughly twice the work of a Large, which is twice a Medium. This predictable arithmetic has guided our planning for years, allowing us to look at past projects and say, ``This feels like a Medium,'' with a high degree of accuracy.

\subsubsection{Why AI Breaks the Ruler}

\textbf{The Exponential Effort Curve.} AI development doesn't follow a straight line; it follows an exponential curve. Improving a model's accuracy from 85\% to 95\% isn't just a small step---it's often a 10x jump in effort, and sometimes it's technically impossible with the current approach. Every percentage point closer to ``perfect'' requires significantly more data, more compute, and more trial-and-error. In AI, the ``last mile'' isn't a mile---it's a marathon. Foundational research on scaling laws demonstrates that model performance follows power-law relationships with compute and data~\cite{kaplan2020,hoffmann2022}, meaning that linear increases in capability require exponential increases in resources.

In agentic systems, these non-linear regimes are measurable. Controlled evaluations show coordination effects can saturate or become negative once single-agent baselines exceed $\sim$45\%, and certain multi-agent topologies can amplify errors substantially~\cite{kim2025}.

Research on agent system scaling~\cite{kim2025} confirms this non-linear behavior, documenting diminishing returns above $\sim$45\% baseline performance and topology-dependent error amplification.

\begin{figure}[htbp]
    \centering
    \includegraphics[width=0.8\textwidth]{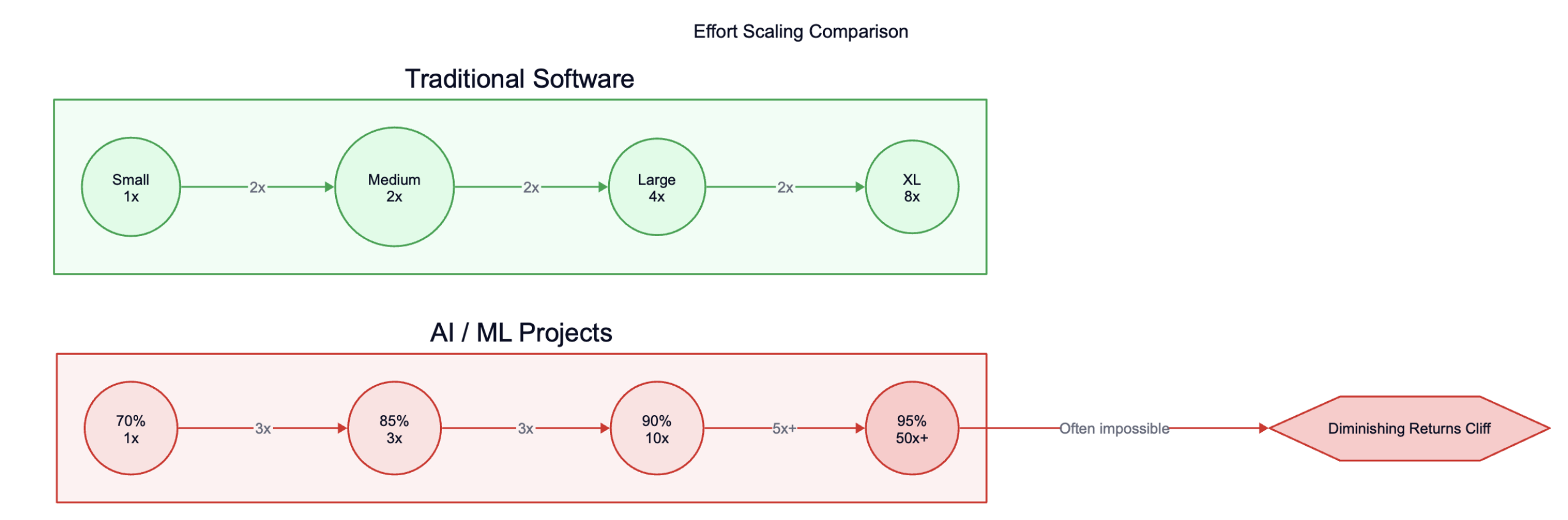}
    \caption{Effort scaling comparison between traditional software (approximately linear) and AI systems (often non-linear).}
    \label{fig:effort-scaling}
\end{figure}

As shown in Figure~\ref{fig:effort-scaling}, in AI, incremental performance gains can require disproportionately greater data, compute, evaluation, and iteration---undermining the ``$2\times$ scope $\rightarrow$ $2\times$ effort'' intuition.

\textbf{Multi-Agent Orchestration Complexity.} Adding agents does not add complexity linearly---it multiplies it. A Supervisor Agent routing tasks across 2 agents faces binary decisions. The same Supervisor across 10 agents faces a combinatorial explosion of routing paths, error states, and emergent failure modes~\cite{cemri2025}.

\begin{table}[htbp]
    \centering
    \caption{Multi-agent interaction complexity grows combinatorially with agent count.}
    \label{tab:agent-complexity}
    \begin{tabular}{@{}lll@{}}
        \toprule
        Agent Count & Required Interaction Tests & Complexity Growth \\
        \midrule
        2  & 2      & Baseline \\
        3  & 6      & +4 \\
        5  & 20     & +14 \\
        10 & 90     & +70 \\
        $N$  & $N(N-1)$ & Combinatorial \\
        \bottomrule
    \end{tabular}
\end{table}

\begin{figure}[htbp]
    \centering
    \includegraphics[width=0.8\textwidth]{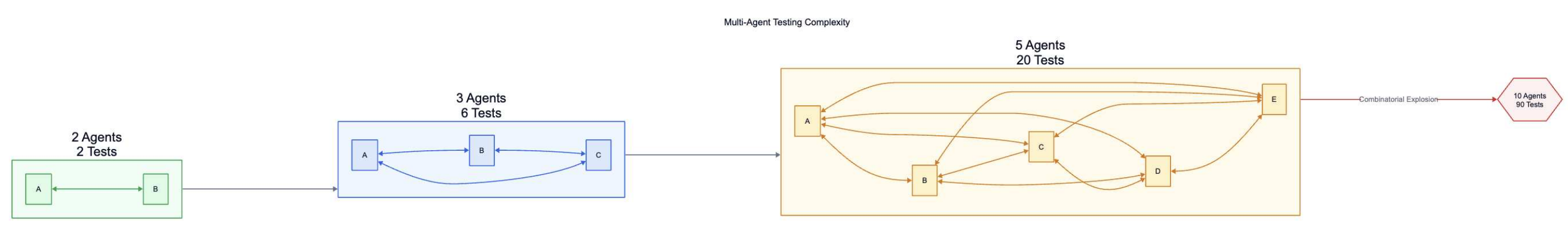}
    \caption{Multi-agent interaction testing complexity grows as $N(N-1)$, where $N$ is the number of agents.}
    \label{fig:interaction-complexity}
\end{figure}

Figure~\ref{fig:interaction-complexity} illustrates why adding a single agent can multiply integration and verification work rather than add a constant increment.

\textbf{Circular Dependency Trap.} Agent A delegates to Agent B, which delegates back to Agent A---creating an infinite loop that silently consumes API budget until detected. Such emergent failure modes are invisible to T-shirt estimates based on component-level sizing.

\begin{figure}[H]
    \centering
    \includegraphics[width=0.8\textwidth]{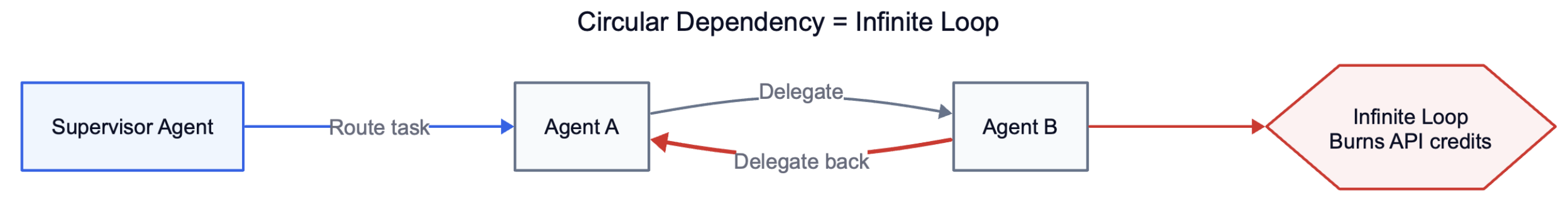}
    \caption{Circular dependency failure mode in multi-agent systems.}
    \label{fig:circular-dependency}
\end{figure}

Interaction dynamics can produce loops (A$\rightarrow$B$\rightarrow$A) that are not visible in component-level estimates as in Figure~\ref{fig:circular-dependency}.

\subsection{Assumption 2: Repeatability from Prior Experience}

\subsubsection{The Traditional Model}

Teams leverage institutional knowledge when estimating. Having built similar APIs or UIs before, they estimate with 70--80\% confidence. Experience maps reliably to future effort. Known unknowns can be buffered; unknown unknowns are rare.

\subsubsection{Why AI Violates This Assumption}

\textbf{Every Dataset is Uncharted Territory.} Building an HR chatbot does not prepare a team to estimate a Technical Support chatbot. Data distributions, vocabulary coverage, edge case density, and annotation quality vary wildly across domains. Teams confront \textbf{Unknown Unknowns}: data corruption surfaces only during training; bias emerges only during evaluation~\cite{bui2025,coelho2024}.

\begin{figure}[H]
    \centering
    \includegraphics[width=0.6\textwidth]{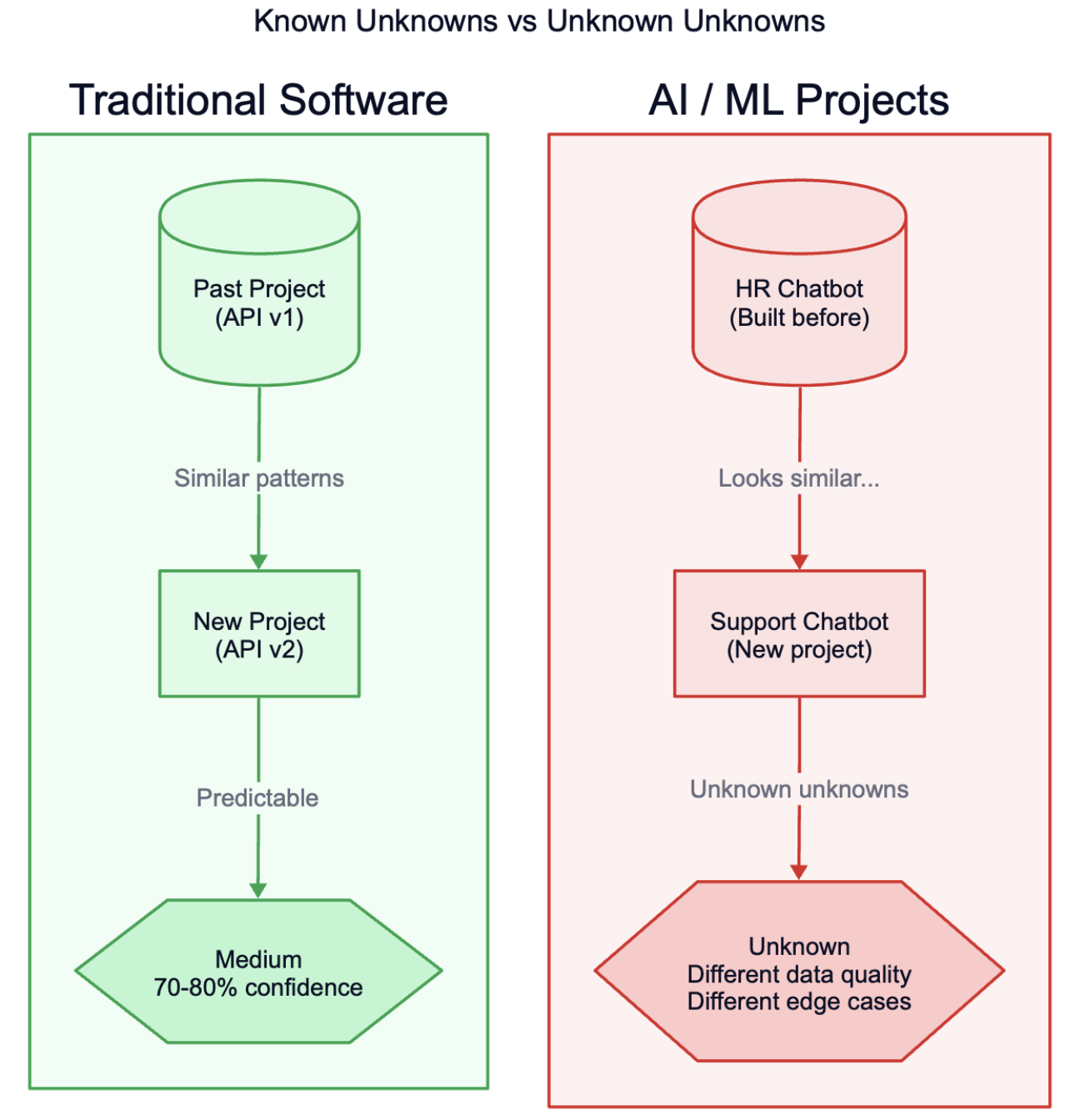}
    \caption{Uncertainty regimes in traditional software vs AI.}
    \label{fig:uncertainty-regimes}
\end{figure}

AI development often contains ``unknown unknowns'' that surface late (e.g., during training, evaluation, or deployment), weakening the value of analogies to prior projects.

\textbf{Context Degradation and State Drift.} Research demonstrates that LLMs exhibit 39\% average performance degradation in multi-turn conversations compared to single-turn interactions~\cite{laban2025}. Over extended multi-turn interactions, agents progressively lose sight of primary objectives, fixating on irrelevant sub-tasks. Context window decay over 20+ turns is unpredictable---hallucinations manifest only in long-duration sessions that were not tested during development.

\begin{figure}[H]
    \centering
    \includegraphics[width=0.8\textwidth]{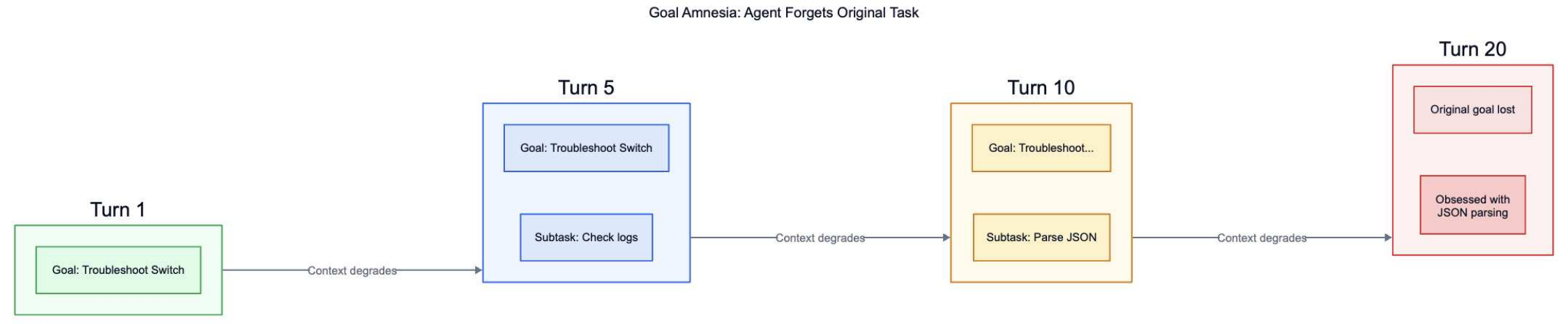}
    \caption{State drift and goal-amnesia in multi-turn LLM interactions.}
    \label{fig:state-drift}
\end{figure}

Large-scale simulations find an average 39\% performance drop from single turn to multi-turn settings and increased unreliability as conversations progress~\cite{laban2025}.

\subsection{Assumption 3: Effort-Duration Fungibility}

\subsubsection{The Traditional Model}

Effort and duration are interchangeable: an 8-week task takes 4 weeks with two engineers. Add resources, compress timelines. Brooks's Law applies at extremes, but modest scaling works for most projects.

\subsubsection{Why AI Violates This Assumption}

\textbf{Irreducible Sequential Dependencies.} AI development contains mandatory sequential phases that cannot be parallelized regardless of team size:

\begin{itemize}
    \item Data collection must precede data cleaning
    \item Model training must precede hyperparameter tuning
    \item Evaluation must precede deployment
\end{itemize}

Each phase has inherent duration floors---compute cycles do not compress with headcount~\cite{kim2025}. Studies of real-world ML deployments document these sequential bottlenecks extensively~\cite{paleyes2022}, while empirical research shows that data work---often the longest phase---is undervalued and cannot be easily parallelized~\cite{sambasivan2021}.

\begin{figure}[H]
    \centering
    \includegraphics[width=0.4\textwidth]{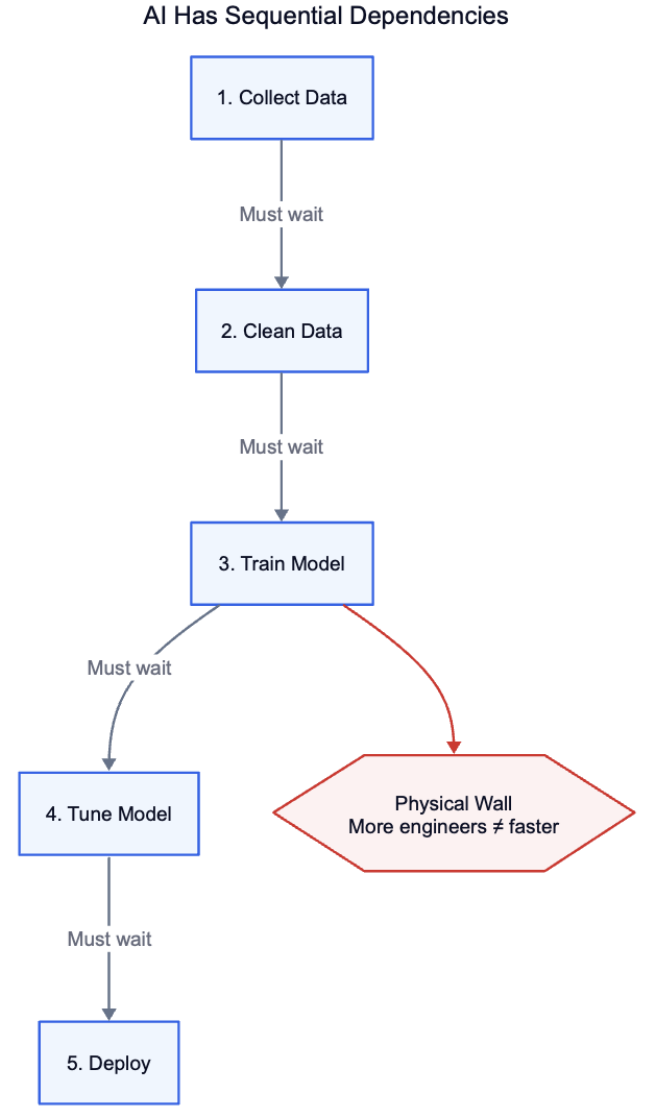}
    \caption{Sequential bottlenecks in AI development pipelines.}
    \label{fig:sequential-bottlenecks}
\end{figure}

As depicted in Figure~\ref{fig:sequential-bottlenecks}, AI projects commonly contain irreducible sequential dependencies (data$\rightarrow$train$\rightarrow$evaluate) that do not compress linearly with added headcount.

\textbf{The Observability Gap.} When a multi-agent workflow fails at step 7, isolating the failure requires replaying steps 1--6 to reconstruct exact memory and conversation state. Traditional debugging approaches fail because agent state is not deterministically reproducible~\cite{cemri2025}.

\textbf{Latency Floors.} Multi-agent pipelines have irreducible latency floors. For example, a 5-agent sequential chain with $\sim$3 seconds per inference call implies a $\sim$15 second minimum end-to-end latency. This can materially constrain UX and throughput and is not captured by category-based estimates.

\begin{figure}[H]
    \centering
    \includegraphics[width=0.8\textwidth]{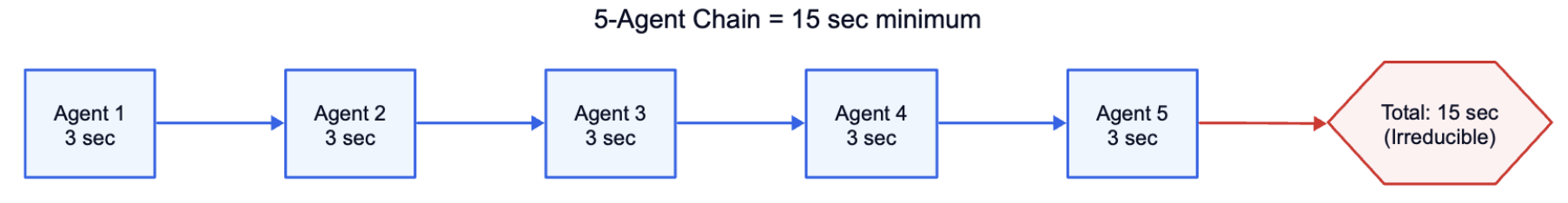}
    \caption{Latency wall in multi-agent pipelines.}
    \label{fig:latency-wall}
\end{figure}

Each agent in a sequential chain contributes inference latency that does not vanish with more engineering effort.

\subsection{Assumption 4: Task Decomposability}

\subsubsection{The Traditional Model}

Large initiatives decompose into parallel workstreams. Four teams tackle four components simultaneously. Integration happens at the end with manageable coordination overhead.

\begin{figure}[H]
    \centering
    \includegraphics[width=0.6\textwidth]{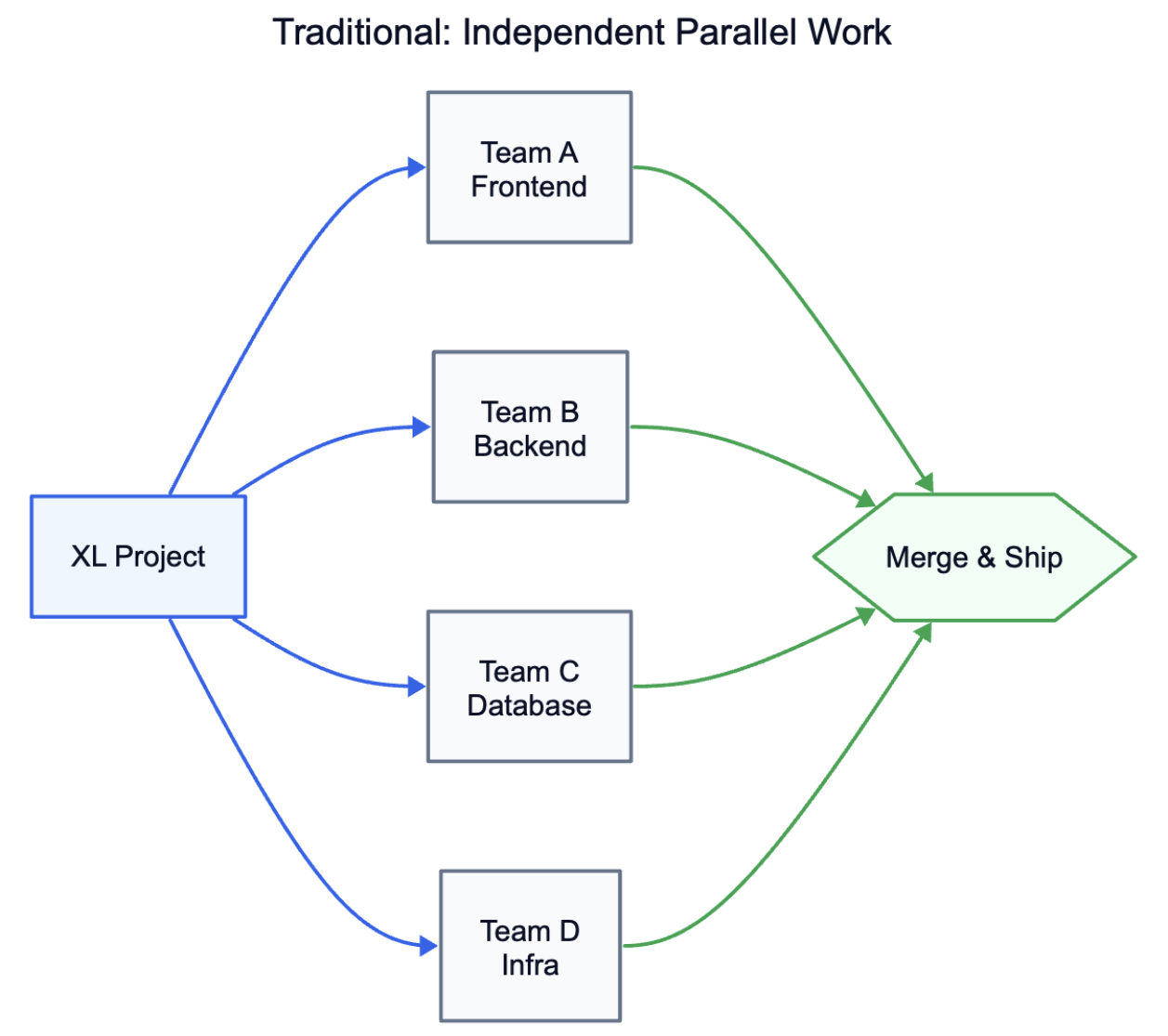}
    \caption{Traditional software decomposition model.}
    \label{fig:decomposition-model}
\end{figure}

Many software systems allow parallel development with stable interfaces, enabling additive estimates across components.

\subsubsection{Why AI Violates This Assumption}

\textbf{Tight Coupling Across the Stack.} Data engineering, model architecture, and prompt engineering form a tightly coupled system. Feature changes in data pipelines cascade into model retraining. Schema changes invalidate prompt logic. Independent T-shirt sizing per component ignores systemic interdependencies~\cite{cemri2025,kim2025}.

\begin{figure}[H]
    \centering
    \includegraphics[width=0.8\textwidth]{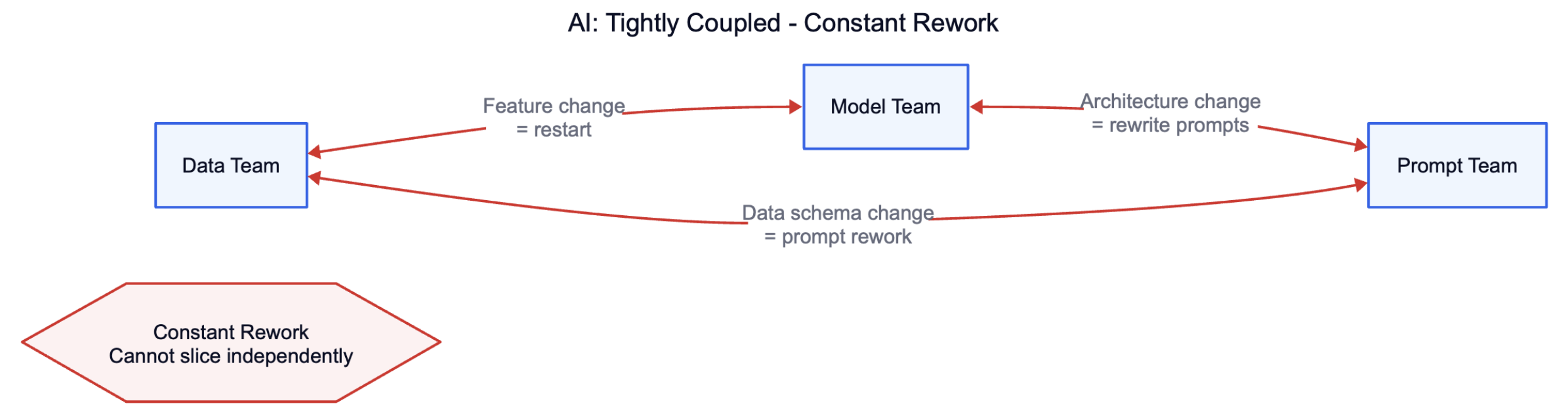}
    \caption{Tight coupling in AI system architecture.}
    \label{fig:tight-coupling}
\end{figure}

Dependencies across data, model, and prompting layers can be bidirectional; changes propagate and trigger rework, breaking additive estimation.

\textbf{Shared Resource Contention.} Multiple agents share a finite context window (e.g., 128k tokens). One verbose agent can exhaust the budget, starving others of reasoning capacity. Global state coupling means Agent B's performance depends on Agent A's output schema---schema evolution breaks downstream reasoning without warning~\cite{laban2025}.

\begin{figure}[H]
    \centering
    \includegraphics[width=0.6\textwidth]{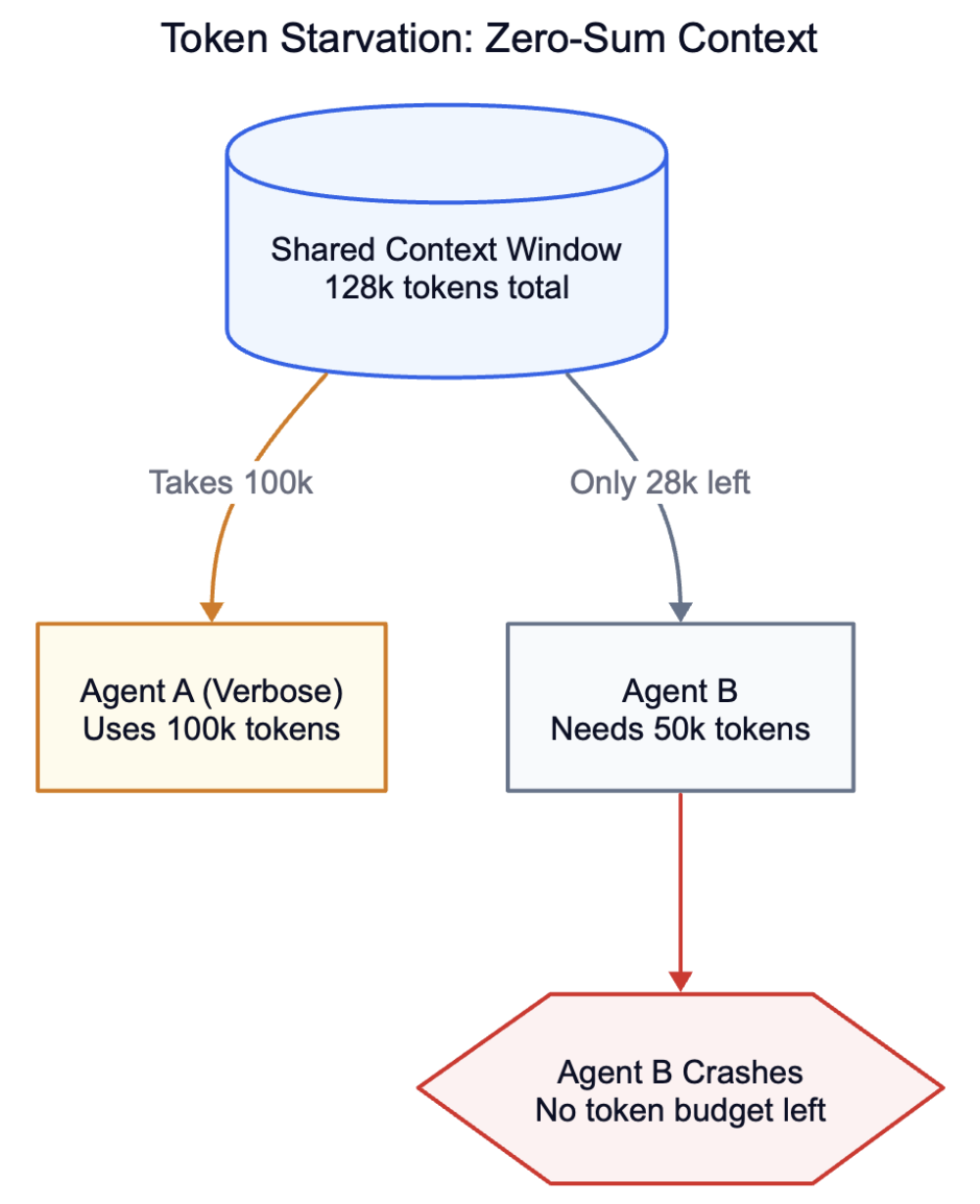}
    \caption{Token starvation in shared context windows.}
    \label{fig:token-starvation}
\end{figure}

Agents compete for finite context capacity; verbose traces can degrade downstream reasoning quality, creating integration risk not captured by component sizing.

\subsection{Assumption 5: Deterministic Completion Criteria}

\subsubsection{The Traditional Model}

Completion criteria are deterministic: tests pass; specifications are met; code ships. The definition of ``done'' remains stable throughout the sprint. A task estimated as Medium stays Medium through completion.

\subsubsection{Why AI Violates This Assumption}

\textbf{The Moving Goalpost Problem.} Accuracy targets are achieved---but Legal rejects the hallucination rate. Performance benchmarks passed---but an ethics audit reveals unacceptable bias. A project declared ``Done'' on Friday resurfaces as an XL on Monday. Non-deterministic failure modes mean completion is probabilistic, not binary~\cite{cemri2025,bui2025}.

\begin{figure}[H]
    \centering
    \includegraphics[width=0.8\textwidth]{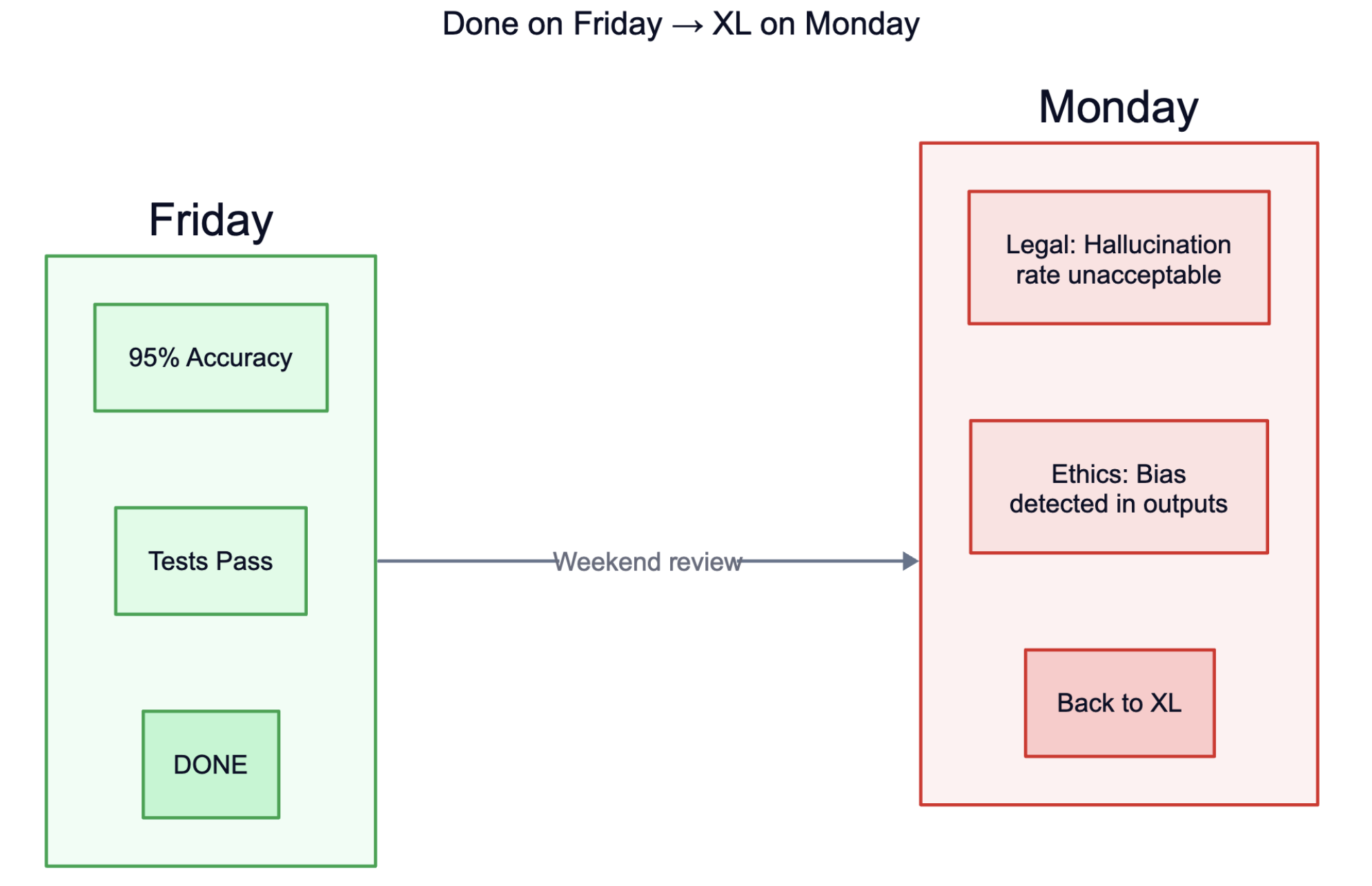}
    \caption{Moving-target problem in AI project completion.}
    \label{fig:moving-target}
\end{figure}

Acceptance criteria can evolve as safety, governance, and reliability constraints are validated---turning an apparently ``done'' task into major remediation.

\textbf{Termination Safety Failures.} Infinite correction loops can emerge when a Supervisor rejects Worker output, the Worker retries, and the Supervisor rejects again---consuming budget without producing progress. Large-scale analyses of multi-agent traces identify recurring failure modes spanning system design issues, inter-agent misalignment, and task verification~\cite{cemri2025}.

\begin{figure}[H]
    \centering
    \includegraphics[width=0.8\textwidth]{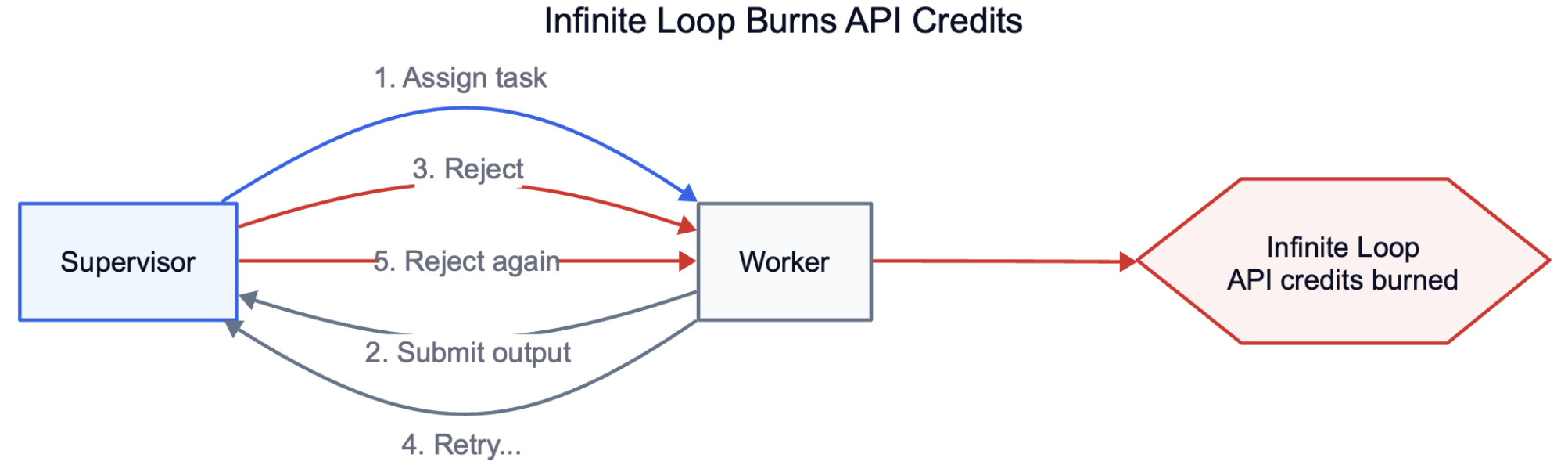}
    \caption{Infinite correction loop failure mode.}
    \label{fig:infinite-loop}
\end{figure}

Without explicit termination and verification strategies, agent systems can enter unproductive retry cycles.

\textbf{Guardrail Oscillation.} Fixing aggressive agent behavior creates passive behavior; loosening constraints creates unsafe behavior. Calibrating the balance requires weeks of RLHF or prompt engineering---a hidden cost invisible to T-shirt estimates.

\begin{figure}[H]
    \centering
    \includegraphics[width=0.8\textwidth]{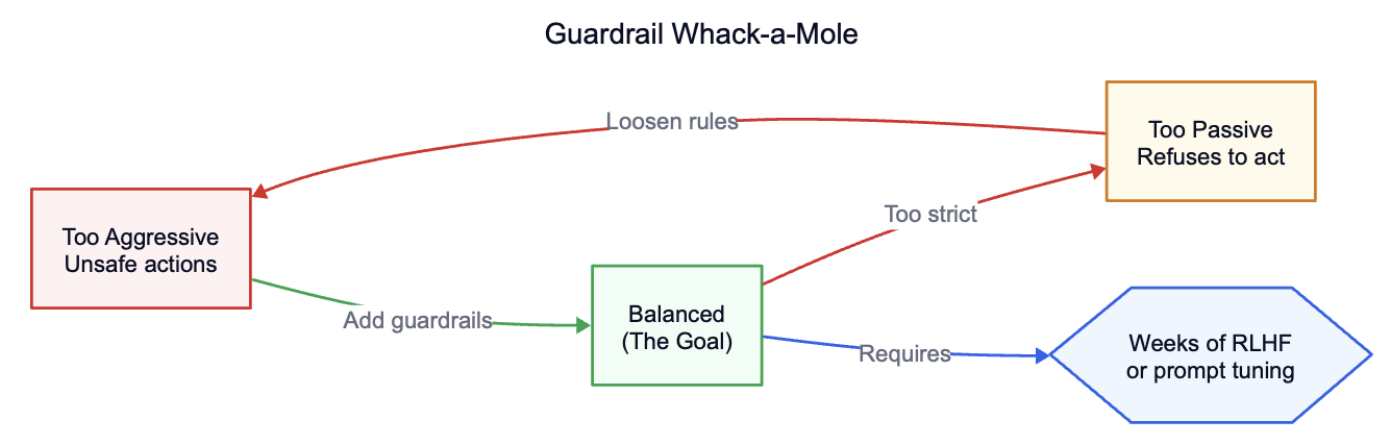}
    \caption{Guardrail oscillation (``whack-a-mole'') in agent behavior tuning.}
    \label{fig:guardrail-oscillation}
\end{figure}

Tightening guardrails can reduce utility; loosening can increase risk. The calibration loop is iterative and can be schedule-dominating.

\section{Discussion}

\subsection{What This Means for Teams}

The analysis in Section 4 shows us that T-shirt sizing doesn't fail because teams are inexperienced or bad at their jobs. It fails because the very foundation of the method---the assumptions we've used for decades---simply doesn't apply to the reality of AI development. If we keep using these old tools, we'll keep seeing the same systemic failures, no matter how much ``buffer'' we add.

Here are the real-world implications:

\begin{enumerate}
    \item \textbf{Buffers Aren't Enough:} You can't just add 20\% to an estimate to cover the ``AI part.'' Exponential curves and complex agent interactions can't be fixed with a simple percentage.
    \item \textbf{Past Performance Isn't a Guarantee:} Just because you built one chatbot doesn't mean the next one will take the same amount of time. Every new dataset brings its own set of ``unknown unknowns.''
    \item \textbf{Estimates cannot be made in a Vacuum:} Because AI layers are so tightly coupled, you can't just size the ``UI'' and the ``Backend'' separately and expect them to sum up to a project total.
    \item \textbf{``Done'' is a Moving Target:} In AI, meeting an initial accuracy goal is often just the beginning. New safety or legal requirements can turn a ``completed'' task back into an XL project overnight.
\end{enumerate}

\subsection{Limitations}

This analysis has several limitations:

\begin{enumerate}
    \item \textbf{Qualitative Evidence:} While grounded in empirical literature, the assumption violations are characterized analytically rather than through a new controlled study conducted by the authors.
    \item \textbf{Generalization:} The analysis focuses on LLM and multi-agent systems; simpler ML projects may violate fewer assumptions.
    \item \textbf{Alternative Methods:} We propose Checkpoint Sizing but do not empirically validate its effectiveness relative to other emerging AI estimation approaches.
\end{enumerate}

\subsection{The Checkpoint Sizing Alternative}

Given the fundamental incompatibility between T-shirt sizing assumptions and AI development characteristics, we propose Checkpoint Sizing as an alternative methodology:

\begin{table}[H]
    \centering
    \caption{Checkpoint Sizing principles and implementation guidelines.}
    \label{tab:checkpoint-sizing}
    \begin{tabular}{@{}p{0.3\textwidth}p{0.6\textwidth}@{}}
        \toprule
        Principle & Implementation \\
        \midrule
        Explicit Decision Gates & Define checkpoints at natural phase boundaries (data readiness, model convergence, safety validation) \\
        Evidence-Based Reassessment & Re-estimate scope, timeline, and feasibility at each checkpoint based on empirical observations \\
        Uncertainty Acknowledgment & Replace point estimates with confidence intervals that reflect genuine uncertainty \\
        Pivot Optionality & Build explicit decision points for scope reduction, pivot, or termination based on checkpoint outcomes \\
        \bottomrule
    \end{tabular}
\end{table}

Checkpoint Sizing treats AI project estimation as an iterative discovery process rather than an upfront planning exercise. Initial estimates serve as hypotheses to be validated through successive checkpoints rather than commitments to be defended.

\subsubsection{A Minimal Procedure (Pseudocode)}

The goal is not ``perfect upfront estimation,'' but \textbf{controlled uncertainty} with explicit decision points.

\begin{algorithm}[H]
\caption{Checkpoint Sizing}
\begin{algorithmic}[1]
\Require InitiativeGoal, InitialHypothesisEstimate, Constraints (budget, latency, safety, timeline)
\Require Checkpoints $= [C_1..C_k]$ (e.g., DataReadiness, EvalHarness, SafetyReliability, CostLatency, Rollout)
\State Estimate $\gets$ InitialHypothesisEstimate
\State Risks $\gets$ initial risk register (unknowns, assumptions, dependencies)
\For{each checkpoint $C_i$ in Checkpoints}
    \State Evidence $\gets$ run checkpoint-specific work and measurements
    \Statex \hspace{\algorithmicindent}\hspace{\algorithmicindent}(data profiling, baseline model tests, eval harness results,
    \Statex \hspace{\algorithmicindent}\hspace{\algorithmicindent}safety probes, cost/latency profiling, red-teaming, etc.)
    \State Update Risks using Evidence (retire risks, add new risks, re-rank severity)
    \State Re-estimate scope/timeline/cost using Evidence (not analogy)
    \State Decision $\gets$ \{Proceed, Pivot, ReduceScope, Pause, Terminate\}
    \If{Decision $\neq$ Proceed}
        \State record rationale and exit (or branch plan accordingly)
    \EndIf
\EndFor
\State \Return final Estimate and delivery plan with residual risks and monitoring hooks
\end{algorithmic}
\end{algorithm}

\subsubsection{Checkpoint Gate Checklist (What You Must Exit With)}

Use these gates as ``definition of readiness'' checkpoints. Each gate produces artifacts and metrics that change the estimate.

\begin{itemize}
    \item \textbf{Gate A --- Data Readiness}
    \begin{itemize}
        \item \textit{Artifacts:} data inventory; schema; quality report; labeling/PII policy; drift risks
        \item \textit{Minimum evidence:} representative samples; missingness/outlier profile; edge-case map; governance sign-off (if required)
    \end{itemize}
    
    \item \textbf{Gate B --- Evaluation Harness}
    \begin{itemize}
        \item \textit{Artifacts:} task taxonomy, gold set, automated eval pipeline, failure categorization
        \item \textit{Minimum evidence:} baseline score + confidence intervals; regression tests; measurement of ``unknown unknowns'' surfaced during eval. Systematic behavioral testing approaches~\cite{ribeiro2020} provide templates for comprehensive evaluation coverage.
    \end{itemize}
    
    \item \textbf{Gate C --- Safety, Reliability, and Control}
    \begin{itemize}
        \item \textit{Artifacts:} safety policy; refusal/grounding strategy; escalation paths; red-team plan; termination/loop controls (agents)
        \item \textit{Minimum evidence:} observed hallucination/error modes; multi-turn reliability characterization~\cite{laban2025}; mitigation plan mapped to failure taxonomy classes~\cite{cemri2025}
    \end{itemize}
    
    \item \textbf{Gate D --- Cost and Latency Budgets}
    \begin{itemize}
        \item \textit{Artifacts:} throughput model; caching strategy; prompt/token budgets; tool-call budget; spend forecast
        \item \textit{Minimum evidence:} measured p50/p95 latency and cost per task; sensitivity to context length and agent count; rollback plan if budgets are exceeded
    \end{itemize}
    
    \item \textbf{Gate E --- Operationalization (Rollout/Monitoring)}
    \begin{itemize}
        \item \textit{Artifacts:} monitoring dashboards; incident playbooks; feedback loops; drift detection; evaluation in production
        \item \textit{Minimum evidence:} canary plan; SLOs; alerting; documented ``definition of done'' that includes post-deploy monitoring criteria
    \end{itemize}
\end{itemize}

\subsection{Illustrative Case Study (Synthetic)}

To show how Checkpoint Sizing changes estimates, consider a synthetic initiative: ``Support Copilot with RAG + Tool Use'' for internal support engineers.

\textbf{Initial T-shirt estimate (Sprint 0):} Large (6--8 weeks), based on analogy to ``build a chatbot'' and prior UI/API delivery.

\textbf{Checkpoint outcomes and estimate evolution:}

\begin{enumerate}
    \item \textbf{Gate A --- Data Readiness}
    \begin{itemize}
        \item \textit{Finding:} 25\% of historical tickets contain redacted fields; knowledge base has inconsistent versions; PII policy requires additional filtering.
        \item \textit{Estimate impact:} +3 weeks (data cleaning + governance).
    \end{itemize}
    
    \item \textbf{Gate B --- Evaluation Harness}
    \begin{itemize}
        \item \textit{Finding:} baseline retrieval fails on long-tail product SKUs; ``answer correctness'' needs rubric + citations; automated eval required for regression.
        \item \textit{Estimate impact:} +2 weeks (eval harness + gold set + metrics).
    \end{itemize}
    
    \item \textbf{Gate C --- Safety, Reliability, and Control}
    \begin{itemize}
        \item \textit{Finding:} multi-turn conversations degrade when users revise constraints; tool calls occasionally produce conflicting states; need loop/termination controls for tool-using agent.
        \item \textit{Estimate impact:} +3--5 weeks (safety probes, guardrails, termination/verification).
    \end{itemize}
    
    \item \textbf{Gate D --- Cost and Latency Budgets}
    \begin{itemize}
        \item \textit{Finding:} sequential tool-using agent path exceeds latency SLO at p95; requires caching + partial parallelization + shorter context strategy.
        \item \textit{Estimate impact:} +2 weeks (performance engineering + architecture changes).
    \end{itemize}
    
    \item \textbf{Gate E --- Operationalization}
    \begin{itemize}
        \item \textit{Finding:} rollout requires monitoring for retrieval drift and hallucination rate; canary plan plus human escalation path required.
        \item \textit{Estimate impact:} +1--2 weeks (observability + rollout controls).
    \end{itemize}
\end{enumerate}

\textbf{Revised estimate after gates:} XL (12--16 weeks) with explicit risk register and exit criteria. The key shift is that the estimate becomes evidence-driven: effort accrues from concrete artifacts (data readiness, eval harness, safety controls, cost/latency budgets, and operationalization) rather than analogy.

\section{Conclusion}

T-shirt sizing was designed for a world of linear scaling, repeatable patterns, effort-duration fungibility, parallelizable work, and deterministic completion. AI development violates all five premises:

\begin{enumerate}
    \item \textbf{Non-linear effort curves} replace linear scaling, with performance gains often requiring disproportionate increases in data, compute, evaluation, and iteration.
    \item \textbf{Dataset uniqueness} eliminates repeatability, making every AI project a journey into terra incognita.
    \item \textbf{Sequential dependencies} break effort-duration fungibility, creating timeline floors that resist resource addition.
    \item \textbf{Tight cross-layer coupling} prevents decomposition, generating cascading rework invisible to component estimates.
    \item \textbf{Probabilistic completion} replaces deterministic criteria, introducing moving goalposts and emergent failure modes.
\end{enumerate}

The path forward requires a new way of thinking, designed specifically for the unique challenges of AI. Checkpoint Sizing is our proposal: a system of explicit decision gates where we reassess scope, timeline, and feasibility based on what we've actually built and learned---not on assumptions inherited from a different era of engineering.

\textbf{Future Work.} Several directions merit further investigation:

\begin{itemize}
    \item \textbf{Empirical Validation:} Controlled studies comparing Checkpoint Sizing against T-shirt sizing and other estimation methods across diverse AI project types.
    \item \textbf{Tooling Integration:} Development of plugins or extensions for project management tools (Jira, Linear, Asana) that operationalize checkpoint gates with templates, artifact checklists, and re-estimation workflows.
    \item \textbf{Domain-Specific Adaptations:} Tailoring checkpoint definitions for specific AI application domains (healthcare, finance, autonomous systems) where regulatory and safety constraints introduce additional estimation complexity.
    \item \textbf{Autonomous Agent Systems:} Extending the analysis to fully autonomous agents with long-horizon planning, where estimation challenges may be even more severe due to open-ended execution and environmental uncertainty.
    \item \textbf{Quantitative Benchmarks:} Establishing benchmark datasets of AI project estimates vs. actuals to enable quantitative comparison of estimation methodologies.
\end{itemize}

\appendix

\newpage
\begin{center}
{\Large\bfseries Appendix}
\end{center}
\addcontentsline{toc}{section}{Appendix}
\vspace{1em}

\subsection*{Reference Validation Methodology}

\subsubsection*{A.1 Validation Process}

Each cited work in the primary references [1]--[5] was evaluated against the five fatal assumptions using a systematic validation process:

\begin{enumerate}
    \item \textbf{Relevance Mapping:} Each paper was mapped to the specific assumptions it directly validates.
    \item \textbf{Evidence Extraction:} Key findings from each paper were extracted and matched to specific claims in Section 4.
    \item \textbf{Coverage Assessment:} The collective coverage of all five assumptions was verified.
\end{enumerate}

\subsubsection*{A.2 Validation Results}

\begin{table}[H]
\centering
\small
\begin{tabular}{@{}p{0.6cm}p{3.2cm}p{6.5cm}p{1.2cm}@{}}
\toprule
\textbf{Ref} & \textbf{Paper Title} & \textbf{Validates Assumptions} & \textbf{Relevance} \\
\midrule
{[}1{]} & Why Do Multi-Agent LLM Systems Fail? & 4 (Decomposability via inter-agent coupling), 5 (Deterministic completion via verification/termination failures) & 100\% \\
\addlinespace
{[}2{]} & An LLM-based multi-agent framework for agile effort estimation & 2 (Repeatability via subjective inconsistency), 5 (Deterministic completion via estimation instability) & 100\% \\
\addlinespace
{[}3{]} & Towards a Science of Scaling Agent Systems & 1 (Non-linear scaling), 3 (Effort-duration tradeoffs), 4 (Coordination overhead/error amplification) & 100\% \\
\addlinespace
{[}4{]} & LLMs Get Lost In Multi-Turn Conversation & 2 (Repeatability via multi-turn unreliability), 5 (Deterministic completion via non-recovering error trajectories) & 100\% \\
\addlinespace
{[}5{]} & Effort and Size Estimation in Software Projects with Large Language Model-based Intelligent Interfaces & 2 (Repeatability via hidden uncertainty), 5 (Definition of done via evolving spec/AI interface behavior) & 100\% \\
\bottomrule
\end{tabular}
\caption{Literature Validation Matrix: Each reference validates specific assumptions with 100\% relevance, providing independent empirical support for the theoretical framework.}
\label{tab:validation}
\end{table}

\subsubsection*{A.3 Assumption Coverage Matrix}

\begin{table}[H]
\centering
\begin{tabular}{@{}lll@{}}
\toprule
\textbf{Assumption} & \textbf{Primary Support} & \textbf{Secondary Support} \\
\midrule
1. Linear Scaling & {[}3{]} & --- \\
2. Repeatability & {[}2{]}, {[}4{]}, {[}5{]} & --- \\
3. Effort-Duration Trade & {[}3{]} & --- \\
4. Decomposability & {[}1{]}, {[}3{]} & --- \\
5. Deterministic Completion & {[}1{]}, {[}2{]}, {[}4{]}, {[}5{]} & --- \\
\bottomrule
\end{tabular}
\caption{Assumption Coverage Matrix: Mapping of literature references to the five fatal assumptions.}
\label{tab:coverage}
\end{table}

All five fatal assumptions receive direct support from empirical sources in [1]--[5], with multiple assumptions supported by more than one work. The 100\% relevance score reflects scope alignment (each reference was selected specifically to substantiate one or more claims in Section 4), not an external meta-analytic quality rating.

\bibliographystyle{unsrt}

\end{document}